# Artificial Intelligence helps making Quality Assurance processes leaner

Alexander Poth[1], Quirin Beck[2], Andreas Riel[3]

[1/2] Volkswagen AG, Berliner Ring 2, D-38436 Wolfsburg, Germany
{quirin.beck|alexander.poth}@volkswagen.de
[3] Grenoble Alps University, F-38031 Grenoble, France
andreas.riel@grenoble-inp.fr

**Abstract.** Lean processes focus on doing only necessery things in an efficient way. Artificial intelligence and Machine Learning offer new opportunities to optimizing processes. The presented approach demonstrates an improvement of the test process by using Machine Learning as a support tool for test management. The scope is the semi-automation of the selection of regression tests. The proposed lean testing process uses Machine Learning as a supporting machine, while keeping the human test manager in charge of the adequate test case selection.

**Keywords:** Artificial Intelligence (AI), Machine Learning (ML), Agile Software Development, Quality Assurance (QA), Testing

## 1  Introduction

Many established long running projects and programs are execute regression tests during the release tests. The regression tests are the part of the release test to ensure that functionality from past releases still works fine in the new release. In many projects, a significant part of these regression tests are not automated and therefore executed manually. Manual tests are expensive and time intensive [1], which is why often only a relevant subset of all possible regression tests are executed in order to safe time and money. Depending on the software process, different approaches can be used to identify the right set of regression tests. The source code file level is a frequent entry point for this identification [2]. Advanced approaches combine different file level methods [3]. To handle black-box tests, methods like [4] or [5] can be used for test case prioritization. To decide which tests can be skipped, a relevance ranking of the tests in a regression test suite is needed. Based on the relevance a test is in or out of the regression test set for a specific release. This decision is a task of the test manager supported by experts. The task can be time-consuming in case of big (often a 4- to 5-digit number) regression test suites because the selection is specific to each release. Trends are going to continuous prioritization [6], which this work wants to support with the presented ML based approach for black box regression test case prioritization.

Any regression test selection is made upon release specific changes. Changes can be new or deleted code based on refactoring or implementation of new features. But also changes on externals systems which are connected by interfaces have to be considered during the tests case selection. This work does not address the methods for how to choose the right indicators for a good selection, as this is considered the job of the test manager. The focus of this work is rather to assist the test manager with a ML based tools which will be trained with the test managers' example selections. Consequently, the tools implicitly applies the test manager's selection criteria in order to come up with a suggestion. After the training, the trained model is applied to the rest of the regression test suite, helping test managers to safe time. The approach is based on the ML based system level test case prioritization [7]

Based on the lean concept principles: value, value stream, flow, pull, perfection [8] the following aspects for the lean regression test case selection can be derived:
- Value: value is generated if the required functionality is validated by only executing necessary tests, which leads to faster time to market and revenue.
- Value stream: the stream of safeguarding a product or service is significantly driven by the selection of the right tests and their right execution order to find failures as early as possible to fix them to optimize the time to market.
- Flow: the selected tests are directly "flying" into the test preparation or execution, the next steps in the testing workflow [9].
- Pull: selected tests are pulled for execution rather than a fixed test suite pushed into the execution pipeline.
- Perfection: the selection process provides iterative and incremental improvement.

Aligned with Toyota Production System (TPS) [10] based plants, implementing these lean principles also implies reducing waste of experts' time for doing things which a machine can do. In this lean context, our investigation topic is therefore to transfer the mechanical automation of the TPS to an ML-based cognitive automation for test expert support during the testing process. Human experts are still required process understanding, execution, improvement and innovation. This is in-line with researchers' expectations that real creativity and innovation we will not see in the next years from ML algorithms [11].

This paper is structured as follows: After a brief investigation of related works in section 2, section 3 proposes an approach to integrating ML algorithms into the existing regression testing process. Section 4 elaborates on the design of a regression test process that is supported by a ML tool learning from and supervised by a human testing expert according to the concept presented in section 3. Section 5 summarizes results of the new process' application in a series of productive projects at the Volkswagen Group. Finally, section 6 concludes with a summary and an outlook.

## 2     Related work

Artificial Neural Networks have been successfully used as white-box test suggestion engines as demonstrated by the survey-articles [12], [13] and [14]. Our approach aims at being independent of the particular ML approach used.

## 3   An approach to integrating ML into regression testing

The objective to support the regression test selection activity with an ML based tool demands a process which is driven by a human as process owner. The responsibility for the decisions is held by the human test manager for the product or service under test. Typically, this other relevant stakeholders having deep knowledge about the release support this expert. The training of the ML model becomes a part of the selection process driven by the test manager, in three major aspects: the test manager
   a) defines the training data set,
   b) defines the verification data set, and
   c) decides about the adequateness of the ML tool suggestion.

The training data is a sub-set of the entire regression test suite (oval in figure 1) and contains examples from the tests which are in (T+ in figure 1) and which are out (T- in figure 1) of the regression test set of the release. Depending on the data quality, less than 100 test cases for each of T+ and T- can be enough for a good training data set.

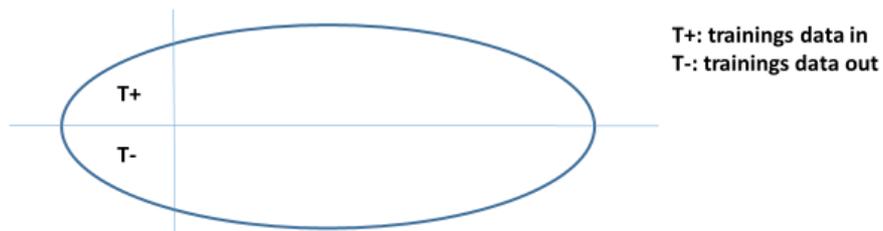

**Fig. 1.** The different data sets to train a regression test set

During the training phase, which can take a few minutes for a big regression test suite, the test manager defines the test data set and selects randomized additional tests (Tv in figure 2) from the regression test suite. The selected tests are be defined as in (green dots in figure 2) or out (red dots in figure 2) for the release regression test.

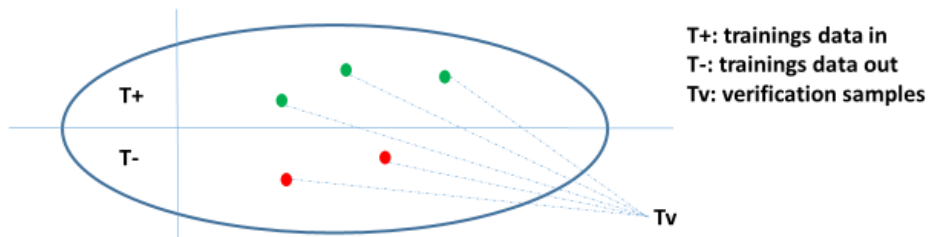

**Fig. 2.** Selection of verification samples for a regression test set

The result of the ML tool is a monotone sequence of tests (blue curve in figure 3). Highly ranked tests in the sequence are in (left of Te in figure 3) and low rank tests are out of the release regression test. The ranking of the tests is compared with the decision of the test manager for the randomized test data set. The dots should be on an adequate position on the curve. Figure 3 shows a good result, because all red and green dots are separated by the decision interval (D). Depending on the result matching of the ML-model and the human test manager, the test manager gets "trust" into the ML based suggestion of the ranking or not. In case of trust or a more formal adequateness check of the ranking, the test manager can define (decide) which test in D is the last test (Te) in the release regression test. The result is inadequate if the green and red dots are mixed (not separated by D). Depending on how much the different colored dots are mixed up, the test manger has to define the ML-suggested ranking as inadequate. In the case of inadequateness, the test manager can improve the result with more trainings data in case of high data quality. However, depending on the data quality, the ML tool may never generate an adequate outcome. The test manager would realize this, and exclude the tool from the selection process until the data quality has changed. The test manager would manually select the test cases. By the process design, all decisions about in/out can still be reused without losing time.

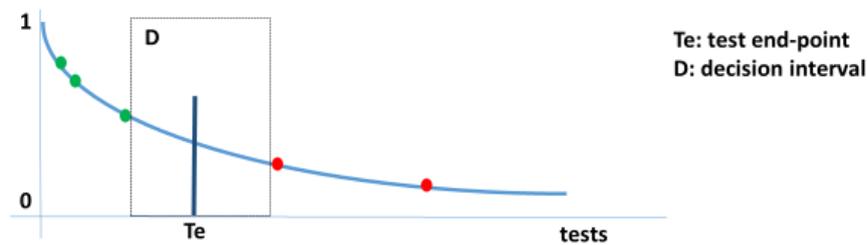

**Fig. 3.** Definition of a regression test set based on ranked tests by the ML based tool

## 4     Regression test process design integrating ML tool support

The process have to be designed to be robust in case of low quality outcomes of the ML-model. The robustness is needed because only a few projects have specialized data scientist capable of evaluating their data, and optimizing tuning their algorithms with hyper-parameters. Integrating ML specialists who can modify an ML algorithm or write a new one for perfect fit is economically infeasible at the current stage. Setting up an environment in which even ML unskilled persons can appropriately work with ML-based tools, implies establishing a process that ensures that low quality result are detected fast and filtered out.

### 4.1 ML tool design for optimal process support

To ensure robustness, the process involves the human test manager on different touch-points with the ML-tool. The interaction on the touch-points have to be clearly defined in terms of what is needed from the human and which ML outputs the latter has to verify. To reach this goal, the ML-tool has been designed in a way as to load the data and give the user the option to de-select some attributes – called features in the ML domain – of these data. This working copy of the selected data contains the requirements, tests, defects and their relations as basis for the training, testing (verification dots in figure 2) and inferencing of the ML algorithm and the for the specific context generated ranking model. The ranking prediction is the outcome of the trained model. The suggested test ranking has to be verified by the human expert.

The issue is how to verify the tool suggestion? This is realized by splitting the data into two data sets [15]. One for training (figure 1 left part with T+ and T-) and another for inferencing (figure 1 right part). The training data can be picked in a randomized way from the entire regression test suite. For an ML algorithm more training data is better than less, however on the saturation point more data will not improve the result significantly. In our evaluation, the saturation point was reached with less than 100 trainings data objects for T+ and T- with a well-structured regression test suite. For optimizing the cycle time, the process is designed to parallelize actions of the human and the machine. During inferencing; the human expert builds the test data set and classifies the tests. The test manager makes some randomized selections of tests from the inferencing data set (figure 2 colored dots) and decides about in/out of the regression test set. To hide complexity from the human expert, the display of the test ranking was considered as not required, leaving only the binary classification results to show. The amount of verification data can be smaller than the training data set, however not too small to avoid accidental results. This verification samples are checked against the suggestion of ML-model (figure 3). D could be smaller than as shown on this simplified example figure in case of T+ and T- "dots" are right or left from the green or red verification dots. This is the crucial point where the test manager has to decide to start an additional iteration with more training data or accept a more or less overlapping. This will necessarily be a case-by-case decision depending mainly on how sure the test manager is about the tolerability of the sample's (mis-)ranking. Based on the answer, a further iteration may start. The iterations can be stopped if more training data do not result in significantly better results/suggestions. If the results do not improve anymore and a significant overlap persists, the test manager will have to stop using the tool (figure 4 show the entire process).

After the definition of this workflow and the responsibilities of the human experts, a data scientist started to analyze different projects from different domains. We selected projects from three different business domains to ensure a wide range of content and different content treatment approaches in order to come up with a generic approach. The different approaches lead to different content structures which the data have to be extracted from for generic processing and feature extraction. The data scientist selected an adequate ML algorithm for processing the data in order to derive the specific trained model. To identify adequate ML algorithms, the state-of-the-art ML frameworks and libraries have been evaluated against the ranking requirements, and the best fitting algorithm selected. Finally, the workflow has been implemented in an easy to use tool with an interface to the enterprise tool suite (API) containing requirements, tests and defects.

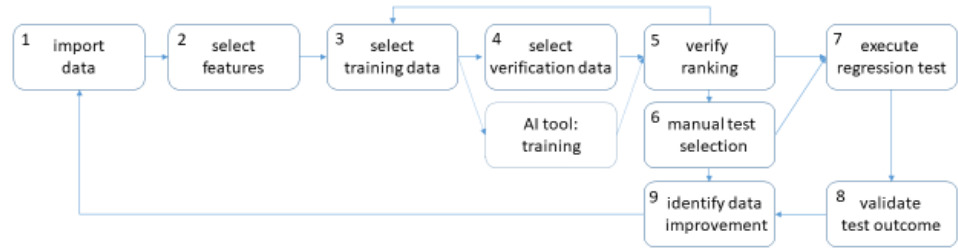

**Fig. 4.** Workflow of the regression test selection for a release.

### 4.2  Rollout kit of the process

To support the lean approach, a rollout strategy was designed to assist agile autonomous teams by mastering the regression test selection. This was realized based on a self-service kit for the projects. To meet this demand, the basic process and its actions (figure 4) have been documented in practical work instruction. The latter is like a recommendation for starting with the tool, integrating be the iterative design the continuous improvement of lean approaches. The process is shown in figure 4. Its first step is to identify the projects data and make them available for the ML-tool. In the second step, the features are scoped for the training. In the third step, a set of trainings data is selected and classified. In the fourth step, the ML model is trained and the test/verification data set defined by the randomized picked tests which are classified by the human. The fifth step checks the adequateness of the outcome of the ML-model. Based on the quality of the outcome the human can decide to accept the suggestion of the ML-model or decide to start another try with more training data (go back to step three). In case of totally inacceptable suggestions the human will realize that the database is not adequate for the ML algorithm and stop working with the tool (step six) and initiate an initiative to identify data improvement (step nine). In a post-test phase step, the human expert shall reflect on the applied regression test set (step eight) and derive points to improve the data base (step nine) which offers more valuable features in the future to be more effective.

In addition to the detailed work instruction, the rollout kit is linked to the tool's code as well as its technical documentation in order to give everybody in the Volkswagen Group IT the chance to analyse any issues in detail. Furthermore, everybody is able to improve the code for better results, in particular data scientists. This openness offers an individual pull of the users because nothing limits their ideas and supports their contributions to the continuous improvement of both the tool and its integration into the process.

## 5   Results and Evaluation

The developed process was evaluated on projects of three different business domains on the basis of APFD [16]. Furthermore, the feature *test case description* was used for the ranking. Our best result was a 0.9336 in the open interval (0,1) prediction of the ML model. The worst prediction was 0.5322. These bad results say that ML based suggestion is only 3.22% better as random picking of tests for the definition of the regression test set, under the assumption that a random prioritization is approximately 0.5. In addition to the test case description, further features were used for the measurement. These could certainly improve the results. The validation was made on past releases of the projects for an easy validation of the outcomes of the ML model with the real world facts of the past release. This high variance is typical for the heterogeneous data and their quality of the different projects. Furthermore, the ML-tool detected a corrupt data set in one of the evaluation projects. This was a positive side-effect observed during the preparation of the feature selection. Depending on the projects, data and the selected training data, the result is more or less satisfying. However, the selection process works fine, taking into account that its quality performance strongly depends on the training know-how of the human experts who selects the training data set. This result fulfills the requirement for ML supporting the human expert. This approach has been made available to all projects and programs in the Volkswagen Group IT to show if the data quality fits with the requirements for training demands of the ML algorithm to have a benefit from the ML-based supporting tool.

## 6   Conclusion and Outlook

This work demonstrates that it is possible to develop a lean process and its tooling which integrates an ML-based support of the daily business of test- and quality managers within a few months only. The current issue is that the hundreds of real life projects of an enterprise have many different approaches to document their requirements, tests and defects, and it is not clear how well the ML algorithm can work – especially learn – from all these data that have not been tuned specifically for ML algorithms. Tuning the data by data scientists is currently not an option for all projects (because of availability and cost of data scientists and efforts for changing established structures of the data and the processes) and the process have to be robust enough to filter out bad recommendations of the ML-model. To handle his unknown data quality the process is designed to give a clear feedback about the outcome quality thanks to mapping of the test samples by the human expert. Therefore, at every time the human can decide about the adequateness of the ML outcomes.

The presented process improvement approach also follows the values and principles that are described in the Software Process Improvement (SPI) Manifesto [17][18][19]. The implementation tries to motivate all the people involved, for example the test manager and other stakeholders who can support the test case selection. It is a dynamic and adaptable way to satisfy customer needs with an agile and lean mindset.